\documentclass[aps,prb,twocolumn,superscriptaddress,floatfix,showpacs]{revtex4}
\usepackage{CJK}
\bibpunct{[}{]}{,}{n}{}{}
\usepackage{graphicx}
\usepackage{dcolumn}
\usepackage{float}
\usepackage{latexsym}
\usepackage{amssymb}
\usepackage{amsfonts}
\usepackage{amsmath}
\usepackage{fancybox}
\usepackage[binary,squaren]{SIunits}
\usepackage{natbib,graphicx,dcolumn,color,lipsum}
\usepackage{subfigure,amsmath,amssymb,verbatim,moreverb,bm,framed}
\usepackage{color}
\usepackage{colordvi}
\usepackage{ulem}

\def\be{\begin{equation}}
\def\ee{\end{equation}}
\def\ber{\begin{eqnarray}}
\def\eer{\end{eqnarray}}

\def\sigmabold{\mbox{\boldmath $\sigma$}}
\def\rv{{\bf r}}

\def\pv{{\bf p}}

\def\vv{{\bf v}}

\def\vv{{\bf v}}
\def\ep{{\epsilon}}

\def\bp{{\bf p}}

\def\nn{\nonumber}
\def\nn{\nonumber}

\def\Acal{{\cal A}}
\def\TSigma{\tilde{ \Sigma}}
\def\TSigmaC{\tilde{\check \Sigma}}
\def\TGC{\tilde{\check G}}
\def\TG{\tilde{ G}}

\def\inten{\int\,\frac{{\rm d}\epsilon}{2\pi i}\,}

\def\Tcal{{\cal T}}
\def\Deltav{{\boldsymbol \Delta}}
\def\Omegav{{\boldsymbol \Omega}}
\newcommand\anticomm[2]{\left\{{#1},{#2}\right\}}

\newcommand{\commentout}[1]{}

\newcommand{\agrave}{\`a}

\DeclareMathAlphabet\mathbfcal{OMS}{cmsy}{b}{n}

\def\be{\begin{equation}}
\def\ee{\end{equation}}
\def\ber{\begin{eqnarray}}
\def\eer{\end{eqnarray}}
\def\nn{\nonumber}
\begin{document}
\begin{CJK*}{GB}{} 

\title{Theory of current-induced spin polarizations in an electron gas}

\author{Cosimo Gorini}
\affiliation{Institut f\"ur Theoretische Physik, Universit\"at Regensburg, 93040 Regensburg, Germany}
\author{Amin Maleki}
\affiliation{CNISM and Dipartimento di Matematica e Fisica, Universit\`a Roma Tre,
Via della Vasca Navale 84, 00146 Rome, Italy}
\author{Ka Shen}
\affiliation{Kavli Institute of NanoScience, Delft University of Technology, Lorentzweg 1, 2628 CJ Delft, The Netherlands}
\author{Ilya V. Tokatly}
\affiliation{Nano-bio Spectroscopy group and ETSF Scientific Development Centre,
Dpto. F\'isica de Materiales, Universidad del Pa\'is Vasco, E-20018 San Sebastian, Spain}
\affiliation{IKERBASQUE, Basque Foundation for Science, 48011 Bilbao, Spain}
\author{Giovanni Vignale}
\affiliation{Department of Physics and Astronomy, University of Missouri,
  Columbia, Missouri 65211, USA}
  \author{Roberto Raimondi}
\affiliation{CNISM and Dipartimento di Matematica e Fisica, Universit\`a Roma Tre,
Via della Vasca Navale 84, 00146 Rome, Italy}



\date{\today}
\begin{abstract}
{  We derive the Bloch equations for the spin dynamics of a two-dimensional electron gas in the presence of spin-orbit coupling. 
For the latter we consider both the intrinsic mechanisms of  structure inversion asymmetry (Rashba) and bulk inversion asymmetry (Dresselhaus),  and the extrinsic ones arising from the scattering from impurities. 
The derivation is based on the  SU(2) gauge-field  formulation of the Rashba-Dresselhaus spin-orbit coupling.   Our main result is the identification of a new spin-generation torque arising from the Elliot-Yafet process, which opposes a similar term arising from the Dyakonov-Perel process. The new spin-generation torque contributes to  the current-induced spin polarization (CISP) -- also known as the Edelstein or inverse spin-galvanic effect.  As a result,  the behavior of the CISP turns out to be  more complex than one would surmise from consideration of the internal Rashba-Dresselhaus fields alone.  In particular, the symmetry of the current-induced spin polarization 
does not necessarily coincide with that of the internal Rashba-Dresselhaus field, and an out-of-plane component of the CISP is generally predicted, as observed in recent experiments. 
We also discuss the extension to the three-dimensional electron gas, which may be relevant for the interpretation  of experiments in thin films.}
\end{abstract}

\maketitle
\end{CJK*}

\section{Introduction}
\label{sec_intro}

Spin polarization by currents and its inverse effect are currently a subject of intensive theoretical and experimental investigation both in semiconducting  and  metallic systems. (For  recent reviews see [\onlinecite{Ganichev2016,Ando2017,Soumyanarayanan2016}]).  Originally the effect was proposed by Ivchenko and Pikus\cite{Ivchenko1978} and observed in tellurium \cite{Vorobev1979}. Later it was theoretically analyzed in a two-dimensional electron gas 
in the presence of Rashba spin-orbit coupling by Edelstein\cite{Edelstein90}, and Aronov and Lyanda-Geller\cite{Aronov89}.
As highlighted in [\onlinecite{Ganichev2016}], the effect, which is the consequence of symmetry-allowed coupling between spin polarization and electrical current, may be referred to under different names.
The generation of a current from a non-equilibrium spin polarization goes under the equivalent names of spin galvanic effect (SGE) or
inverse Edelstein effect (IEE), while the reciprocal phenomenon of current-induced spin polarizations is referred to as 
inverse spin galvanic effect (ISGE) or Edelstein effect (EE).

The first experimental observation of the SGE/IEE was in quantum wells, 
by measuring the current produced by the absorption of polarized light \cite{Ganichev2001,Ganichev02,Ganichev2006}.
 More recently it has been shown that a non-equilibrium spin polarization injected by spin-pumping into an Ag$|$Bi \cite{Sanchez13} 
or Fe$|$GaAs \cite{chen2016} interface also yields an electrical current.  Notice that spin-to-charge conversion in this context 
is sometimes referred to as inverse Rashba-Edelstein effect~\cite{ShenVR14} -- 
for a recent theoretical discussion see \cite{toelle2016}.   The SGE/IEE has also been observed at ferromagnet-topological insulator interfaces \cite{Mellnik2014,Shiomi2014} 
and in LAO$|$STO systems \cite{Lesne2016}.
In semiconducting structures the reciprocal ISGE/EE is measured via optical detection of the current-induced spin polarization 
\cite{Kato04,Sih2005,Yang2006,Chang2007,Norman2014}.  The ISGE/EE is also measured by analyzing the torques exercised, via  exchange coupling, by the non-equilibrium polarization on the magnetization of an adjacent ferromagnetic system\cite{Kohno2006,Miron2011,Tatara2013,Amin2016,Amin2016b,Emori2016,Ghosh2017}. Recently this has been extended also to antiferromagnets\cite{Zelezny2017}.

Although the key mechanism of the effect relies on the symmetry properties of gyrotropic media\footnote{See Ref.~\onlinecite{Ganichev2016}. A system is called gyrotropic
if in its point symmetry group some components of polar vectors and
components of axial vectors transform according to the same
representation.}, most of the recent theoretical work has concentrated on models based on the Rashba and Dresselhaus spin-orbit coupling
(respectively RSOC and DSOC in the following) in the presence of disorder scattering responsible 
for spin relaxation\cite{Vavilov2005,Raimondi06,Engel2007,Tarasenko07,Trushin2007,Raichev2007,
Milletari2008,Raimondi09,GoriniPRB10,Golub2011,Malshukov2011,Raimondi_AnnPhys12,Golub2013,ShenVR14,Shenprb14,Inglot2014,Inglot2015,Malshukov2016}.  
In a 2-dimensional electron gas, as for instance the one studied in Ref.~\cite{Norman2014}, spin-orbit coupling (SOC) 
is also due to the electric field of the scattering impurities and the interplay between intrinsic and extrinsic mechanisms becomes highly non trivial\cite{Raimondi09,Raimondi10,Raimondi_AnnPhys12}.
In this paper we analyze some aspects of this interplay focusing on the description of the of ISGE/EE and SGE/IEE 
in a 2-dimensional electron gas -- we will however also discuss results valid in 3d gases.

The model Hamiltonian reads
\begin{eqnarray}
H&=&\frac{p^2}{2m}+\alpha p_y\sigma_x-\alpha p_x\sigma_y+\beta p_x\sigma_x-\beta p_y\sigma_y \nonumber\\
&+&V(\rv)-\frac{\lambda_0^2}{4}\sigmabold\times \nabla V(\rv)\cdot \pv,\label{ham}
\end{eqnarray}
where $\pv$, $\rv$ and $\sigmabold$ represent the momentum, coordinate and spin (in units of $\hbar /2$) operators, respectively, for electrons in the 2-dimensional electron gas. 
The latter lies in the $(x,y)$ plane, while $m$ is the effective mass and $\alpha, \beta$ are the RSOC and DSOC coupling constants.
$V(\rv)$ is a random potential describing the scattering from the impurities.  The potential is assumed to have   zero average and second moment given by $\langle V(\rv)V(\rv')\rangle=n_iv_0^2\delta(\rv-\rv')$, where $v_0$ is the scattering amplitude from a single impurity and $n_i$ is the impurity concentration.  Finally, $\lambda_0$ is the effective Compton wavelength describing the strength of the extrinsic spin-orbit coupling.

 The layout of the paper is as follows: 
The core message is summarized in Sec.~\ref{sec_statement}, where we immediately ``get into the thick of it'' by formulating the problem
and providing its solution.  We base our discussion on physical arguments only, and postpone the technical details
substantiating our conclusions to Secs.~\ref{sec_SU2} - \ref{sec_SU2_ext}, which can thus  be skipped by
the reader not interested in the specifics of our derivation.
More precisely, in Sec.~\ref{sec_SU2} we briefly review the SU(2) approach employed to build the kinetic equation 
in the presence of the RSOC and DSOC, and in Sec.~\ref{sec_bloch_int} we derive the Bloch equations 
when only intrinsic SOC is present.  Finally, in  Sec.~\ref{sec_SU2_ext} we present a rigorous derivation of the  Bloch equations, including the corrections arising from extrinsic effects.   
Here we obtain the crucial new spin-generation torque arising from the Elliot-Yafet process, and discuss its implications for the  ISGE/EE and SGE/IEE
in specific experimental setups.
Finer details concerning the calculation
of the collision integral are provided in Appendices  \ref{appb} and \ref{appa}.
We adopt units such that $\hbar=c=1$ throughout.

\section{The problem and its solution}
\label{sec_statement}

Consider an ensemble of carriers in a generic solid state environment,
where the spin is not a conserved quantity.  In a homogeneous sample,
in the presence of an exchange/Zeeman field $\Deltav$, the ensemble spin polarization ${\bf S}$  
will then obey the continuity (Bloch) equation
\be
\label{bloch1}
\partial_t S^a = - [{\Deltav\times{\bf S}}]^a + \Tcal^a
\ee
where here and throughout Latin superscritps stand for spin components $a=x,y,z$.
The first term on the r.h.s. describes precession around the field $\Deltav$,
while $\Tcal^a$ is the $a$-th component of the torque acting on the spin polarization,
responsible for relaxation to equilibrium.  In a simple isotropic medium it reads 
\be
\label{torque1}
\Tcal^a = -\frac{1}{\tau_s}\left(S^a - S^a_{\rm eq}\right),
\ee
where $\tau_s$ is the spin relaxation time -- of whatever origin -- and the equilibrium spin density 
${\bf S}_{\rm eq} = \chi\Deltav$ is given in terms of the Pauli spin susceptibility $\chi=\frac{1}{4}\partial n/\partial\mu$ which at zero temperature reduces to $\chi=N_0/2$, 
with $N_0$ the density of states per spin at the Fermi energy.

In the presence of intrinsic SOC a finite drift velocity ${\bf v}$ of the ensemble
is associated with a non-equilibrium spin polarization.  Let us take a drift $v_x$ in the $x$-direction
and RSOC for definiteness' sake.  The non-equilibrium spin polarization then reads
\be
\label{ISGE1}
S^y = \chi B^y,
\ee
with
\be
\label{SO_field1}
B^y = 2m\alpha v_x
\ee
an effective ``drift field" felt by the moving ensemble. 
When the drift is caused by an electric field, $v_x = -e\tau E_x/m$,
Eqs.~\eqref{ISGE1} and \eqref{SO_field1} describe the usual  ISGE/EE.\cite{Edelstein90, Aronov89}
RSOC also leads to (anisotropic) Dyakonov-Perel spin relaxation
\be
1/\tau_s \rightarrow 
\hat{\Gamma}_{DP} = \frac{1}{\tau_{DP}}
\left(
\begin{array}{ccc}
1 & 0 & 0 \\
0 & 1 & 0 \\
0 & 0 & 2
\end{array}
\right),
\label{new_eq_6}
\ee
with $1/\tau_{DP} = (2m\alpha)^2D$ and $D=v^2_F\tau/2$ the diffusion constant.
This suggests that we modify the Bloch equations \eqref{bloch1} to
\be
\label{bloch2}
\partial_t S^a = - [\Omegav\times{\bf S}]^a + \Tcal^a_{int},
\ee
where $\Omegav=\Deltav + {\bf B}$ is the full effective 
exchange/Zeeman field felt by the drifting carriers and (repeated indices are summed over, unless otherwise specified)
\be
\label{torque2}
\Tcal^a_{int} = -\hat{\Gamma}_{DP}^{ab}\left(S^b - \chi \Omega^b\right)
\ee
is the intrinsic torque, ``intrinsic"  meaning that spin-orbit effects from impurities are not yet included.
This torque has a spin-relaxation component,
$-\hat{\Gamma}_{DP}{\bf S}$, and a spin-generation one, $\hat{\Gamma}_{DP}\chi\Omegav$.  
The intuitive form of Eqs.~\eqref{bloch2} and ~\eqref{torque2} will be rigorously justified  in Sec. \ref{sec_bloch_int},
and holds for any kind of intrinsic SOC -- e.g. RSOC + DSOC -- with the appropriate form of
$\hat{\Gamma}_{DP}$ and $\Omegav$. 
It shows that the spin polarization relaxes to a non-equilibrium steady-state value given by
\be
{\bf S}_{\rm neq} \equiv \chi\Omegav={\bf S}_{\rm eq} + \chi {\bf B}\,. 
\ee

What happens to this intuitive picture once extrinsic SOC is taken into account?  
This is the central {\it problem} addressed in our work. 
While modifications to both the relaxation and the spin generation torques are clearly expected,
their precise form is a priori far from obvious.  
This is because extrinsic SOC gives rise to several phenomena,
such as side-jump, skew scattering, and Elliott-Yafet relaxation, which are not necessarily
additive with respect to intrinsic SOC effects.\cite{Raimondi_AnnPhys12}
Let us start with the spin relaxation torque, which acquires a contribution due to Elliott-Yafet scattering
\be
\label{relax_torque}
\hat{\Gamma}_{DP}\,{\bf S} 
\rightarrow [\hat{\Gamma}_{DP} + \hat{\Gamma}_{EY}]\,{\bf S} 
\equiv \hat{\Gamma}\,{\bf S},
\ee
with $\hat{\Gamma}_{EY} \sim \lambda_0^4$.  Unsurprisingly, spin-flip events at impurities, 
which are second order in the extrinsic SOC constant $\lambda_0^2$, 
provide a parallel channel for relaxation. 
However they also crucially affect the non-equilibrium steady-state value ${\bf S}_{\rm neq}$  the spins want to relax to.
This is subtler, and highlights the difference between a true equilibrium state and
a non-equilibrium steady-state.  Such state is determined by the spin generation torque, which extrinsic SOC modifies in two ways:  
First via side-jump and skew scattering, which together add an extrinsic contribution $\theta_{ext}^{sH} \sim \lambda_0^2$ to the intrinsic spin Hall angle, $\theta_{int}^{sH} \sim (\alpha^2, \beta^2)$ { (this can have the same or the opposite sign as the intrinsic angle)}
\footnote{The precise form of $\theta^{sH}_{extr}$ depends on various microscopic details, but is inconsequential for our discussion and thus will not be specified. Explicit microscopic expressions, however, can be found in Ref. \cite{Raimondi_AnnPhys12}.}.
Second, via Elliott-Yafet relaxation, { which yields a correction {\it opposite} to the non equilibrium part of the intrinsic spin generation
term, i.e., the $\hat{\Gamma}\chi B$    part of $\hat{\Gamma}\chi \Omega$  in Eq.~\eqref{torque2}}:
\ber
\label{gen_torque}
\hat{\Gamma}_{DP}\chi\Omegav &\rightarrow& 
\left[\hat{\Gamma}_{DP} + \hat{\Gamma}_{EY}\right]\,\chi\Deltav + 
\nn\\
&&
\quad\quad
+ \left[\hat{\Gamma}_{DP}+\hat{\Gamma}_{DP} \frac{\theta_{ext}^{sH}}{\theta_{int}^{sH}} - \hat{\Gamma}_{EY}\right]\,\chi{\bf B}
\nn\\
&\equiv&
\hat \Gamma\,{\bf S}_{\rm eq}+ \delta\hat{\Gamma}\,\chi{\bf B}\,,
\eer
where
\be\label{Gamma}
\hat \Gamma= \hat{\Gamma}_{DP}+\hat{\Gamma}_{EY}\,,
\ee
and
\be\label{DeltaGamma}
\delta \hat \Gamma= \hat{\Gamma}_{DP}+\hat{\Gamma}_{DP} \frac{\theta_{ext}^{sH}}{\theta_{int}^{sH}}- \hat{\Gamma}_{EY}\,.
\ee

The full Bloch equations thus become
\be
\label{bloch3}
\partial_t S^a = -[\Omegav\times{\bf S}]^a -\hat{\Gamma}^{ab} (S^b - \chi \Delta^b)
+\delta\hat{\Gamma}^{ab}\chi B^b.
\ee
 This is the  main result of our paper.  It shows that, while intrinsic and extrinsic SOC act in parallel
as far as relaxation to the equilibrium state is concerned -- second term on the r.h.s. of Eq.~\eqref{bloch3} --
they compete for the more interesting non-equilibrium contribution -- { the spin-generation torque, described by the third term on the r.h.s. of Eq.~\eqref{bloch3}}.  In particular, the last term on the right hand side of Eq.~(\ref{DeltaGamma}) describes an ``Elliot-Yafet spin-generation torque", which {\it opposes} the more familiar Dyakonov-Perel and spin Hall terms.   Physically, this can be understood as follows.  
The non-equilibrium spin polarization appears when the drifted Fermi surface is split by intrinsic SOC, 
the outer surface having a larger spin content than the inner one.  Such an effect is counteracted by
Elliott-Yafet relaxation, which is proportional to the square of the momenta before and after scattering,
and therefore more  efficient for states on the outer surface and less efficient for states on the inner surface.

Eq.~\eqref{bloch3} shows that the naive Bloch equation~\eqref{torque2} is modified by extrinsic processes.  While this fact had already been recognized in previous works 
(Refs.~\cite{Raimondi_AnnPhys12, ShenVR14}) some terms (third order in SOC: first order in RSOC and second order in $\lambda_0^2$) 
of the diagrammatic expansion had been neglected  leading to an incomplete form of  $\delta\hat{\Gamma}$, 
in which the last term on the right hand side of Eq.~\eqref{DeltaGamma} was missing.  
As a result, the numerical calculation of current-induced spin polarization must be reconsidered.

The non-trivial modification of the ISGE/EE arising from Eq.~\eqref{bloch3} implies a corresponding 
modification of the SGE/IEE, so as to fullfill Onsager relations.
To be explicit, in the scenario reciprocal to the one considered in Eqs.~\eqref{ISGE1}, \eqref{SO_field1},
the charge current $J_x$ generated by a non-equilibrium spin polarization $S^y-\chi\Delta^y$
acquires the correction
\be
\label{onsager}
\delta J_x = \frac{2e\alpha\tau}{\tau_{EY}}\left(S^y-\chi\Delta^y\right).
\ee
This ensures reciprocity between the spin response to an electric field $E_x$ and the charge response
to a time-dependent magnetic field $-\Delta^y(t)$ \cite{ShenVR14}.
A microscopic derivation of \eqref{onsager} in a more general context is discussed in Ref.~\cite{toelle2016}
and will not be pursued here.

\section{The SU(2) approach for intrinsic SOC}
\label{sec_SU2}

A convenient way to deal with the RSOC and DSOC of Hamiltonian of Eq.(\ref{ham}) 
is the SU(2) approach, where the SOC is described in terms of a spin-dependent gauge field \cite{GoriniPRB10}.   
This formalism, introduced in the context of quark-gluon kinetic theory \cite{ElzHei1989,WeiHei1991}, 
was recently also extended to superconducting structures with SOC \cite{BerTok2013PRL,BerTok2014PRB}.
For a recent pedagogical introduction, see Ref.~\cite{Raimondi2016}. 
Here we limit ourselves to recall the key aspects of the approach to make this presentation self-contained.
Neglecting for the time being the extrinsic SOC, the RSOC and DSOC of (\ref{ham}) can be written in the form of a spin-dependent vector potential and the Hamiltonian reads
\begin{equation}
\label{ham2}
H=\frac{({\bf p}+e {\mathbfcal A}^a\sigma^a /2)^2}{2m}-\frac{e \Psi^a\sigma^a}{2 }+ V(\rv ),
\end{equation}
where terms $\mathcal{O}(\mathbfcal{A}^2)$ are dropped, as they are second-order in
${\mathcal A}/p_F \ll1$ \footnote{Furthermore, for homogeneous spin-orbit fields,
as in the present case, such terms give rise to a constant offset which can be absorbed in the chemical potential.}.
The only non zero components of ${\mathbfcal  A}^a$ are
\begin{equation}
\label{vecpot}
e {\cal A}_x^x=2m\beta,  \
e {\cal A}_x^y=-2m\alpha,  \
e {\cal A}_y^x=2m\alpha,  \
e {\cal A}_y^y=-2m\beta.
\end{equation}
Relations (\ref{vecpot}) follow by comparing (\ref{ham2}) with (\ref{ham}). In the Hamiltonian we have also included a Zeeman term
\begin{equation}
\label{zeeman}
H_Z=-\frac{\Delta^a\sigma^a}{2}\equiv -\frac{e \Psi^a\sigma^a}{2},
\end{equation}
which can be seen as a spin-dependent scalar potential. 
In the above $\Delta =g_L \mu_B  B_{exter}$ with $g_L$ the gyromagnetic factor, $\mu_B$ the Bohr magneton and $B_{exter}$ the external magnetic field.
In this way the theory can be written in terms of a SU(2) gauge theory of electrons coupled to a $d$-potential gauge field $( \Psi,  {\mathbfcal A})$, where each component of the $d$-vector is expanded in Pauli matrices. Notice that in this description the standard scalar and vector potentials can be included as  the identiy $\sigma^0$ components. 
For the sake of generality, we formulate the theory in $d$ dimensions. Whereas our first motivation is the description of the spin dynamics in a 2DEG, our conclusions apply also to the three-dimensional electron gas. This is specially relevant in experimental situations where one deals with semiconducting thin films.
In the following we make use of the compact (relativistic)  space-time notations for the potentials
\be
{\cal A}^{\mu}=(\Psi, {\mathbfcal A}),  \  {\cal A}_{\mu}=(-\Psi, {\mathbfcal A}),
\label{eq_9_a}
\ee
 the coordinate and momentum
\be
x^{\mu}=(t,\rv), \ x_{\mu}=(-t,\rv), \ p^{\mu}=(\ep,\pv), \ p_{\mu}=(-\ep,\pv)
\label{eq_9}
\ee
and the corresponding derivatives
\be
\partial^{\mu}\equiv \frac{\partial}{\partial x_{\mu}}, \
\partial_{\mu}\equiv \frac{\partial}{\partial x^{\mu}}, \
\partial_p^{\mu}\equiv \frac{\partial}{\partial p_{\mu}}, \
\partial_{p,\mu}\equiv \frac{\partial}{\partial p^{\mu}}.
\label{eq_10}
\ee
In this way the product $p^{\mu}x_{\mu}=-\ep t +  \pv\cdot \rv$ has the correct Lorentz metrics.  We also introduce mixed Wigner coordinates given by the center-of-mass coordinates $(t,\rv)$ and energy-momentum variables $(\epsilon, \pv)$, which are the Fourier-transformed variables of the relative coordinates.

According to the analysis of [\onlinecite{GoriniPRB10}], a semiclassical Boltzmann kinetic equation can be derived from a microscopic  Keldysh formulation in the presence of non-Abelian gauge fields. The starting point is the left-right subtracted 
Dyson equation
\be
\left[\check G_0^{-1}(x_1,x_3)\overset{\otimes}{,}\check G(x_3,x_2) \right]=\left[\check \Sigma(x_1,x_3)\overset{\otimes}{,}
\check G(x_3,x_2) \right],
\label{eq_3}
\ee
where we have used space-time coordinates $x_1\equiv (t_1,\rv_1)$ {\it etc}, 
and quantities with a ``check'' 
($\check G_0^{-1}, \check G, \check \Sigma$) are matrices in Keldysh space \cite{rammer1986}. 
In Eq.~\eqref{eq_3} the symbol $\otimes$ implies integration over $x_3$ and matrix multiplication both in Keldysh and spin spaces.
Furthermore
\be
 \check G_0^{-1}(x_1,x_3)
=\left( i\partial_{t_1}-H \right)\delta (x_1-x_3),
\label{eq_4}
\ee
where $H$ is the Hamiltonian operator (\ref{ham2}),
while the self-energy $\check \Sigma$ appearing in the collision kernel [right hand side of \eqref{eq_3}]
will be specified later.
 
The key step, with respect to the standard way of obtaining semiclassical transport theories {\agrave} la Boltzmann from their microscopic counterparts, is the introduction of a locally covariant Green function $\check {\tilde G}(x_1,x_2)$ (to be defined in the following).  From the Wigner transformed covariant Green function $\check {\tilde G}(p,x)$ one can define the SU(2) covariant distribution function to be determined by the kinetic equation.
The introduction of the covariant Green function in the presence of non-Abelian gauge fields generalizes the well known shift in the momentum dependence of the Green function when one wants to make it gauge invariant under U(1) gauge transformations\cite{Langreth1966,Altshuler1978}. In the SU(2) case, as shown in Ref.~\cite{GoriniPRB10}, such a shift, due to the non commutative nature of the symmetry group, can be carried out in terms of Wilson lines of the gauge field, whose definition is recalled in Appendix \ref{appb}. For our purposes, under the assumption that the spin-orbit energy scale is small compared to the Fermi energy, it is enough to perform the shift to lowest order in the 
gauge field and, as shown in Appendix \ref{appb}, obtain 
\be
\label{shift1}
\tilde{\Phi}(p,x)=\Phi(p,x)-\frac{1}{2}\left\{e{\cal A}^{\mu},\partial_{\mu,p}\Phi(p,x)\right\},
\ee
where $\Phi(p,x)$ is any quantity in the Wigner representation to which the shift can be applied. The
inverse transformation reads
\be
\label{shift2}
{\Phi}(p,x)=\tilde{\Phi}(p,x)+\frac{1}{2}\left\{e{\cal A}^{\mu},\partial_{\mu,p}{\tilde \Phi}(p,x)\right\}.
\ee
We stress that in obtaining Eqs.(\ref{shift1}-\ref{shift2}) terms $\it{O}({\cal A}^{\mu}\partial_{p,\mu})^2$ have been neglected in the above, and will be throughout the paper.

In order to obtain the SU(2) Boltzmann equation from the quantum kinetic equation we 
apply the transformation (\ref{shift1}) to the Eq.(\ref{eq_3}) and to the matrix Keldysh Green function 
\begin{equation}
{\check G}=
\left(\begin{array}{cc}G^R & G^K \\0 & G^A\end{array}\right)
\rightarrow \check {\tilde G}=
\left(\begin{array}{cc}\TG^R & \TG^K \\0 & \TG^A\end{array}\right),
\end{equation}
where $G^{R,A,K}$ denotes respectively the retarded, advanced and Keldysh Green's function \cite{rammer1986}.
As a result we get
\ber
\TG^R-\TG^A& =& -2\pi i\delta(\epsilon-\epsilon_{\bp}),\quad\label{cov_gr}\\
\TG^K &=& -2\pi i\delta(\epsilon-\epsilon_{\bp})\left[1-2f(\pv,x)\right],\label{cov_gk}
\eer
where $\epsilon_{\pv}=p^2/2m -\mu$ measures the energy with respect to the  the chemical potential $\mu$.
Notice that the SU(2)-shifted spectral density ($\sim \TG^R-\TG^A$) has no spin structure: the latter is all in the distribution function $f(\pv,x)$. The fact that the locally-covariant ${\tilde G}^{R,A}$ do not depend on the
gauge fields, i.e. on the RSOC and DSOC, is the great advantage of the approach as will appear later.

Finally, the equation for $\check {\tilde G}$ reads
\be
V^{\mu} \left[\tilde\partial_{\mu}\check{\tilde G}+\frac{1}{2}\Big\{e{\cal F}_{\mu\nu},\partial_{p}^\nu\check{\tilde G}\Big\} \right]=\left[\check{\tilde \Sigma},\check {\tilde G}\right], \label{eq_27}
\ee
where $V^{\mu}=(1, \pv/m)$ is the $d$-current operator  and we have introduced the covariant derivative
\be
\tilde\partial_{\mu}\check{\tilde G}=\partial_{\mu}\check{\tilde G} + i\left[e {\cal A}_{\mu}, \check{\tilde G}\right]
\label{eq_28}
\ee
and the field strength
\be
 {\cal F}_{\mu\nu}=\partial_{\mu}{\cal A}_{\nu}-\partial_{\nu}{\cal A}_{\mu} +
i e \left[ {\cal A}_{\mu},{\cal A}_{\nu}\right].
\label{eq_29}
\ee
An intuitive way to understand Eq.(\ref{eq_27}) is  by noticing that the combination $V^{\mu}\partial_{\mu}$ is the ordinary hydrodynamical derivative
entering the Boltzmann equation,
$\partial_t +\vv \cdot \nabla_{\rv}$, written in compact $d$-vector notation. Furthermore, in the case of the Abelian U(1) electromagnetic gauge field,
the combination $V^{\mu} F_{\mu\nu}\partial^{\nu}_p$ yields the familiar Lorentz force. Eq.(\ref{eq_27}) represents its 
extension to the SU(2) scenario, as will become clear in the following.
The right hand side of Eq.(\ref{eq_27}) follows by applying the covariant transformation to the 
Keldysh collision kernel, $I_K=-i\left[\Sigma,G\right]$ and taking advantage of the unitarity of the Wilson line
as shown in Appendix \ref{appb} (cf. Eq.(\ref{appb_6})).
By taking the Keldysh component of \eqref{eq_27} and separating time and space components we get
\be
\left(\tilde\partial_t +\frac{\pv}{m}\cdot \tilde \nabla_{\rv}\right)f(\pv, \rv, t)
-
\frac{1}{2}\Big\{ {\bf F}\cdot \nabla_{\pv}, f(\pv, \rv, t)\Big\}=I
\label{eq_34}
\ee
 where $\bf F$ is the spin-dependent force due to the SU(2) gauge fields
\begin{equation}
 \label{force}
F_i = e{\cal F}_{0i} + e\frac{p_k}{m}{\cal F}_{ki} = e{\cal E}_i  +e\epsilon_{ikj}\frac{p_k}{m}{\cal B}_j.
\end{equation}
Here ${\cal E}_i={\cal F}_{0i}$ and ${\cal B}_i=\frac{1}{2}\epsilon_{ijk}{\cal F}_{jk}$ are the SU(2) electric and magnetic fields, respectively.
 
\begin{figure}
\includegraphics[width=3.in]{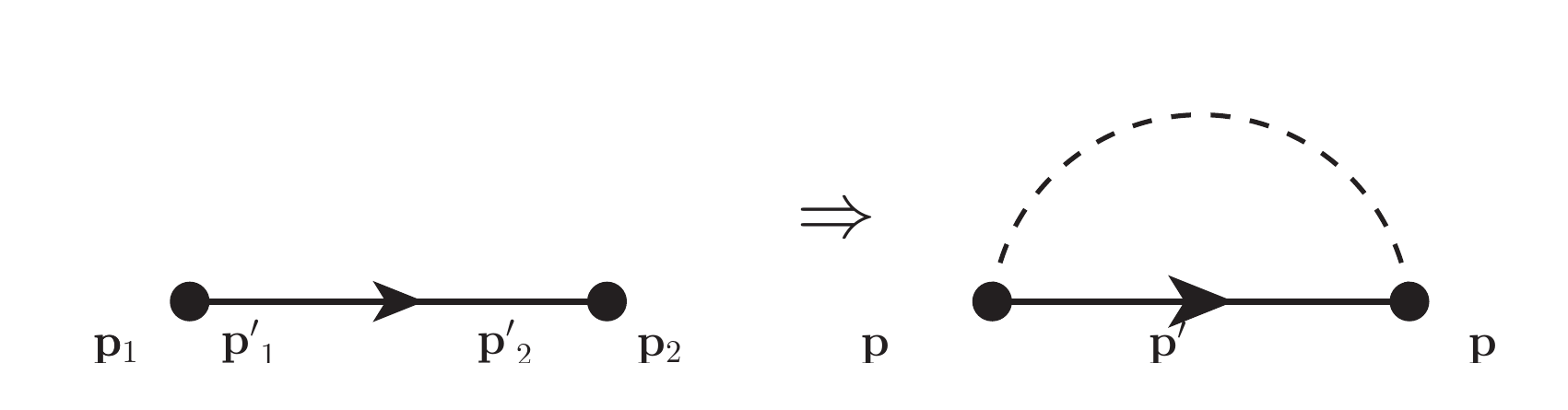}
\caption{Self-energy diagram to second order in the impurity potential (black dot vertex). The diagram on the left is before the impurity average, which is carried in the diagram on the  right  as a dashed line connecting the two impurity insertions. }
\label{fig1}
\end{figure}
The Boltzmann collision integral in Eq.~(\ref{eq_34}) is given by
\be
I\equiv \int \frac{{\rm d}\epsilon}{2\pi i} I_K =-\int \frac{{\rm d}\epsilon}{2\pi}\left[ \check{\tilde \Sigma}, \check{\tilde G}_{\pv}\right]^K.
\label{eq_42}
\ee
To the level of the self-consistent Born approximation, the self-energy due to the disorder potential  is shown in Fig.\ref{fig1} and reads
\footnote{Since the scattering is elastic $p^{\mu}=(\epsilon, \pv)$ and $p'^{\mu}=(\epsilon, \pv')$, i.e. the energy does not change.}
\be
\check \Sigma_0 (p,x)=n_{i}v_0^2\sum_{\pv'} \check G(p',x)
\label{eq_39}
\ee
and yields the familiar Boltzmann collision integral for impurity scattering
\be
I[f]=-2\pi n_i v_0^2\sum_{\pv'} \delta (\ep_{\pv}-\ep_{\pv'}) (f(\pv,\rv,t)-f(\pv',\rv,t)).
\label{eq_45}
\ee
In the next Section we show how the Bloch equations can be obtained starting from the generalized Boltzmann equation (\ref{eq_34}).

\section{The ``intrinsic'' Bloch equations}
\label{sec_bloch_int}

As shown in [\onlinecite{GoriniPRB10}], the spin density  and spin current density defined by 
\ber
S^a (\rv, t)&=&\sum_{\pv} {\rm Tr} \left[ f(\pv, \rv, t)\frac{\sigma^a}{2}\right], \nn\\
{J}^a_i(\rv,t)&=&\sum_{\pv}\frac{p_i }{m} {\rm Tr} \left[ f(\pv, \rv, t)\frac{\sigma^a}{2}\right],
\label{eq_35}
\eer
obey a continuity-like equation
\begin{equation}
\label{continuity}
\tilde\partial_t S^a +\tilde\partial_{i} {J}^a_i=0.
\end{equation}
The above result can be derived from Eq.(\ref{eq_34}) by taking the Pauli matrix component $\sigma^a$ and integrating over the momentum.
After making explicit the covariant derivatives according to (\ref{eq_28}),  the continuity-like equation (\ref{continuity}) becomes
\begin{equation}
\label{continuityexp}
\partial_t S^a +\epsilon_{abc} e\Psi^b S^c+\nabla_i J_i^a-\epsilon_{abc} e{\cal A}_i^b J^c_i=0.
\end{equation}
Here $\epsilon_{abc}$ is the fully antisymmetric Ricci tensor.
The second term in Eq.(\ref{continuityexp}) is the standard {\it precession} term due to the Zeeman term (\ref{zeeman}). The last term of (\ref{continuityexp}) can be made explicit by providing the expression for the spin current $J_i^a$, where the lower (upper) index indicates the space (spin) component. 
In [\onlinecite{GoriniPRB10}] (cf. therein Eq.(68)) the expression of $J^a_i$ was derived via a microscopic theory in the diffusive regime. The expression reads 
\begin{equation}
\label{spincurrent}
J_i^a=v_iS^a -D(\nabla_iS^a -\epsilon_{abc}e {\cal A}_i^b S^c)
-\frac{e\tau n}{4m}({\cal E}^a_i + \epsilon_{ijk}v_j{\cal B}^a_k)
\end{equation}
where $v_i=-\frac{e\tau}{m}E_i$ is the average drift velocity of electrons driven by the external electric field. All the terms in Eq.~(\ref{spincurrent}) have a specific physical origin. The first is a {\it drift} term, containing the spin density $S^a$ carried by the electrons drifted by the electric field $E_i$. The second  is a {\it diffusion} term that contains two contributions: (i) the standard diffusion current proportional to $\nabla_iS^a$, and (ii) the contribution originating from the gauge-field part of the covariant derivative (\ref{eq_28}) acting on the spin density.    
The third term corresponds the {\it SU(2) drift} current driven by the spin-dependent force of Eq.~(\ref{force}). In particular the second contribution in this term yields the  spin {\it Hall} coupling due to the SU(2) magnetic field ${\cal B}^a_i$.

Because of non Abelian nature of the SU(2) gauge group the corresponding magnetic and electric fields can be nonzero even for spatially homogeneous potentials provided their components are not commuting. In this important special case the SU(2) magnetic and electric fields are determined by the commutator term in Eq.~(\ref{eq_29}) (cf. also Eqs.(25-30) in [\onlinecite{GoriniPRB10}])
\begin{eqnarray}
\label{spinmagnetic}
 \epsilon_{ijk}{\cal B}^a_k &=&-\epsilon_{abc}e{\cal A}_i^b {\cal A}_j^c, \\
\label{spinelectric}
{\cal E}^a_i &=& \epsilon_{abc}e{\cal A}_i^b\Psi^c. 
\end{eqnarray}
Using this representation for the fields and recalling the Einstein relation $\frac{\tau n}{m}=D\frac{\partial n}{\partial\mu}\equiv 4D\chi$ one can combine the gauge potential-dependent terms in Eq.~\eqref{spincurrent} into a single item, and rewrite the expression for the spin current in the following compact form
\begin{equation}
\label{spincurrent2}
J_i^a=v_iS^a -D\nabla_iS^a + D\epsilon_{abc}e {\cal A}_i^b\left( S^c - \chi\Omega^c\right)
\end{equation}
where ${\bm \Omega}$ is the total magnetic field introduced in Sec.~\ref{sec_statement}:
\begin{equation}
 \label{Omega}
\Omega^a = e\Psi^a - e{\cal A}_k^a v_k \equiv \Delta^a + B^a.
\end{equation}
Here ${\bm \Delta}$ is the usual Zeeman field defined after Eq.~(\ref{zeeman}) and ${\bf B}$  represents the internal SOC field induced by the electric current (electric field)
\begin{equation}
 \label{Bgeneral}
B^a = -e{\cal A}_k^a v_k = \frac{e\tau}{m}e{\cal A}_k^a E_k.
\end{equation}

Now the Bloch equation describing the global spin dynamics in the presence of intrinsic SOC can be derived by assuming a homogeneous spin density ($\nabla_i{\bf S}=0 $) and substituting the spin current of Eq.~(\ref{spincurrent2}) into Eq.~(\ref{continuityexp}). The resulting equation reads
\begin{equation}
\label{bloch}
\partial_t {S}^a=-({\boldsymbol\Omega}\times {\bf S})^a- \hat \Gamma_{DP}^{ab} \left({S}^b-\chi {\Omega}^b\right),
\end{equation}
where ${\Omega}^a$ is given by Eq.~(\ref{Omega}) and the DP relaxation tensor $\hat \Gamma_{DP}^{ab}$ is defined as follows
\begin{equation}
 \label{DPtensor}
\hat \Gamma_{DP}^{ab} = e^2D\epsilon_{afc}\epsilon_{bdc}{\cal A}_i^d{\cal A}_i^f = 
e^2 D\left(\delta^{ab}{\cal A}_i^c{\cal A}_i^c - {\cal A}_i^a{\cal A}_i^b\right).
\end{equation}
Equation (\ref{bloch}) generalizes Eq.(\ref{bloch2}) of Section \ref{sec_statement} to the case of arbitrary intrinsic SOC. It is worth noticing that in the present formalism  the DP relaxation arises as the {\it second order}  covariant derivative (the covariant Laplacian). One needs to act twice with the gauge field to get the quadratic dependence on the SOC in the spin relaxation matrix.

The second term on the right hand side of the Bloch equation \eqref{bloch} corresponds to the intrinsic torque 
${\cal T}^a_{int}$ for generic SOC. The part of ${\cal T}^a_{int}$ proportional to the internal SO field ${\bf B}$ (\ref{Bgeneral}) can be recognized as the spin generation torque
\begin{equation}
 \label{sg-torque}
{\cal T}^a_{int,sg} = \hat \Gamma_{DP}^{ab} B^b = e^4D\frac{\tau}{m}{\cal A}_i^b({\cal A}_i^b{\cal A}_k^a -
{\cal A}_k^b{\cal A}_i^a)E_k.
\end{equation}
The intrinsic spin generation torque ${\cal T}^a_{int,sg}$ is given by the covariant  divergence of the spin Hall current, that is the very last term proportional to ${\cal B}_k$ in Eq.~(\ref{spincurrent}). Therefore the spin generation torque vanishes for the configurations of the gauge potentials with vanishing SU(2) magnetic field. These configurations correspond to a so called pure gauge SOC for which different space components of the SU(2) potential are commuting and the intrinsic spin Hall effect is absent. Our results imply that in this situation the current-induced spin polarization is also absent.

It is instructive to write explicitly the above general formulas for the specific form of the vector potential of Eq.(\ref{vecpot}) which corresponds to the Rashba-Dresselhaus SOC. In this case the  SU(2) magnetic field has only one nonzero component
\begin{equation}
\label{spinhallfield}
e{\cal B}^z_z=-e{\cal F}_{xy}^a=(2m\beta )^2 -(2m\alpha )^2.
\end{equation}
As the SU(2) magnetic field determines the spin Hall coupling it can be expressed in terms of the 
spin Hall angle for the intrinsic SOC defined by
\begin{equation}
\label{shaint}
\theta_{SH}^{int} =m\tau (\beta^2-\alpha^2)=\frac{e\tau{\cal B}^z_z}{4m}.
\end{equation}
The expression for the spin Hall angle has a suggestive interpretation by recalling the classical Hall effect where
the coupling between the mutually orthogonal charge currents is given by the product of the cyclotron frequency and the scattering time $\omega_c \tau =e B_{exter}\tau/m$. In the present case to get the spin Hall angle
(\ref{shaint}) one needs to combine the SU(2) cyclotron frequency $e{\cal B}^z_z/(4m)$ with the scattering time $\tau$.
An intuitive way to understand the origin of the factor of 4 in the denominator of the SU(2) cyclotron frequency is the following.
Let us imagine that spin up and spin down particles undergo the ordinary Hall effect in opposite directions 
with a spin-dependent magnetic field,
$j_y^{\uparrow}=(\tau /m) B^{\uparrow}j_x^{\uparrow}$ and $j_y^{\downarrow}=-(\tau /m) B^{\downarrow}j_x^{\downarrow}$.
By defining the spin current as $J^z_y=(j_y^{\uparrow}-j_y^{\downarrow})/2$ and identifying $B^{\uparrow}=-B^{\downarrow}={\cal B}^z_z/2$, one
immediately finds the "SU(2)" cyclotron frequency $e{\cal B}^z_z/(4m)$.

By introducing further  an in-plane Zeeman field $e \Psi^x\equiv \Delta^x$ and $e\Psi^y\equiv \Delta^y$, we find that the only nonzero components of the SU(2) electric field are
\begin{eqnarray}
e {\cal E}^z_x &=& \Delta _x 2m \alpha +\Delta_y 2 m\beta \label{zeeman1}\\
e {\cal E}^z_y &=& \Delta _x 2m \beta +\Delta_y 2 m\alpha.
\label{zeeman2}
\end{eqnarray}
In this case the total magnetic field ${\boldsymbol \Omega}$ of Eq.~(\ref{Omega}) also has only in-plane components
\begin{eqnarray}
\Omega^x &=&\Delta^x +B^x\label{effmagx}\\
\Omega^y &=& \Delta^y+B^y\label{effmagy}
\end{eqnarray}
with the internal SO field ${\bf B}$ (\ref{Bgeneral}) of the form
\begin{eqnarray}
B^x&=&2e\tau (\beta E_x +\alpha E_y)\label{int_field_x}\\
B^y&=&-2e\tau  (\alpha E_x +\beta E_y).\label{int_field_y}
\end{eqnarray}

The general DP relaxation matrix $\hat \Gamma_{DP}$ of Eq.~(\ref{DPtensor}) entering Eq.(\ref{bloch}) simplifies as follows
\begin{equation}
\label{DPmatrix}
\hat\Gamma_{DP} =
\begin{pmatrix}
  \tau_{\alpha}^{-1}+\tau^{-1}_{\beta}    &   2\tau^{-1}_{\alpha\beta}&0 \\
   2\tau^{-1}_{\alpha\beta}   &   \tau_{\alpha}^{-1}+\tau^{-1}_{\beta}&0\\
   0 & 0& 2( \tau_{\alpha}^{-1}+\tau^{-1}_{\beta} )
\end{pmatrix}
\end{equation}
where $\tau^{-1}_{\alpha}=(2m\alpha)^2 D$, $\tau^{-1}_{\beta}=(2m\beta)^2 D$ and
$\tau^{-1}_{\alpha\beta}=(2m)^2 \alpha\beta  D$. Notice that for $\beta =0$  the matrix $\hat \Gamma_{DP}$ becomes diagonal, and $\tau_{\alpha}$ reduces to the Dyakonov-Perel relaxation time introduced in Eq.(\ref{new_eq_6}).
Finally, the spin generation torque reads
\ber
\label{spingeneration}
{\mathbfcal  T}_{int, sg}& \equiv& \hat \Gamma_{DP} \chi {\bf B}\\
& =&-2m \theta^{int}_{SH}
(-2e N_0 D)
\begin{pmatrix}
  -\alpha E_y +\beta E_x        \\
     -\beta E_y +\alpha E_x \\
     0   
\end{pmatrix}.\nn
\eer
The above equation generalizes the spin generation torque introduced in Eq.(\ref{torque2}) to the case of  RSOC and DSOC for arbitrary direction of the electric field. \footnote{Notice that ${\mathbfcal T}_{int, sg}$ corresponds to $\gamma$ in the notations of Ref. [\onlinecite{Norman2014}].}  In agreement with the general discussion after Eq.~(\ref{sg-torque}) the spin generation torque is proportional to the spin Hall angle. Therefore it vanishes for SOC giving $\theta^{int}_{SH}=0$ which in the present case corresponds to the compensated RSOC and DSOC with $\alpha=\pm\beta$.

The meaning of Eq.~(\ref{bloch}) is that, under stationary conditions, ${\bf S}=\chi {\boldsymbol \Omega}$,  provided the spin Hall angle is nonzero. This implies that the spin polarization follows the total magnetic field and  (for an energy-independent scattering time~\cite{GoriniPRB10})  there can be no out-of-plane spin polarization since ${\boldsymbol \Omega}$ lays in the xy plane.
This is no longer the case when one considers the extrinsic SOC as will be shown in the following Section.

\section{The effects of extrinsic SOC}
\label{sec_SU2_ext}

The interplay of intrinsic and extrinsic SOC was investigated previously in \cite{Raimondi09,Raimondi10,Shenprb14,Raimondi2016}.
According to the analysis therein Eq.~\eqref{bloch} acquires two modifications.  The first, to order $\lambda_0^2$, is an additional contribution to the spin Hall coupling in the third term in expression (\ref{spincurrent}) for the spin current.
This arises from the inclusion of side-jump and skew-scattering effects due to the extrinsic SOC  and
 leads to a renormalization of the spin Hall angle in the expression of the spin generation torque in Eq.(\ref{spingeneration})
\begin{equation}
\label{sharen}
\theta^{int}_{SH}\rightarrow \theta_{SH}=\theta^{int}_{SH}
+\theta^{ext}_{SH}.
\end{equation}
The second term, which arises to order $\lambda_0^4$,  is an additional contribution to the spin relaxation matrix (the EY spin relaxation). In fact, as discussed in Section \ref{sec_statement}, there exists, to the same order $\lambda_0^4$,  a third {\it new} contribution, which will be derived
 in detail in the following.
\begin{figure}
\begin{center}
\includegraphics[width=2.in]{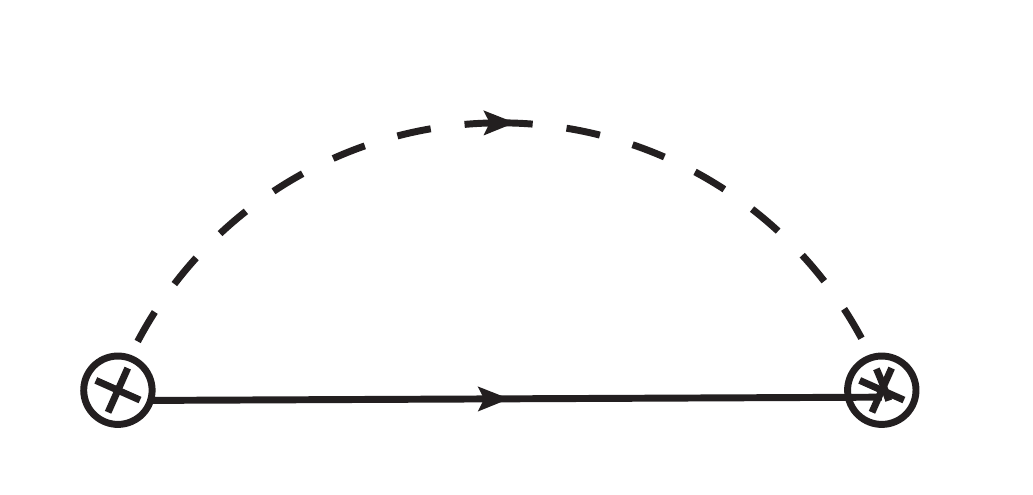}
\caption{Self-energy diagram  in second order in the spin-orbit impurity potential, shown as a crossed empty dot,  contributing to the Elliott-Yafet spin relaxation.}
\label{fig4}
\end{center}
\end{figure}

To see how the new contribution arises, we focus on the term of order $\lambda_0^4$ in the self-energy, whose Feynman diagram is shown in Fig. \ref{fig4} and whose expression reads 
\begin{eqnarray}
\label{born}
{\check \Sigma}_{EY}(\pv) &=&n_i\sum_{\pv'}{\hat V}_{\pv,\pv'} {\check G}_{\pv'}{\hat V}_{\pv',\pv},\label{selfenergy}
\end{eqnarray}
where ${\hat V}_{\pv,\pv'}$ is the spin-dependent part  of  the impurity scattering amplitude
\begin{equation}
{\hat V}_{\pv,\pv'}=iv_0({\lambda_0}/{2})^2(\pv\times\pv')\cdot\sigmabold.
\end{equation}
Shifting  the self-energy of Eq.(\ref{selfenergy}) according to SU(2) shifts (\ref{shift1}-\ref{shift2}) yields the locally covariant EY self-energy  
\ber
{\tilde{ \check {\Sigma} }}_{EY} =
{\tilde{ \check {\Sigma} }}_{EY}^{(0)} +{\tilde{ \check {\Sigma} }}_{EY}^{(1)}.
\label{covariantEY}
\eer
In Eq.(\ref{covariantEY})  we separated the term responsible for the EY relaxation
\ber
{\tilde{ \check {\Sigma} }}_{EY}^{(0)} = n_i\sum_{\pv'}{\hat V}_{\pv,\pv'} {\tilde{\check G}}_{\pv'}{\hat V}_{\pv',\pv}
\eer
from that giving rise to the new contribution
\ber
{\tilde{ \check {\Sigma} }}_{EY}^{(1)}=\frac{n_i}{2}\sum_{\pv'}&(&{\hat V}_{\pv,\pv'} \{\Acal_k,\partial_{p'_k}{\tilde{\check G}}_{\pv'}\}{\hat V}_{\pv',\pv}\\
&-&\{\Acal_k, \partial_{p_k}{\hat V}_{\pv,\pv'} {\tilde{\check G}}_{\pv'}{\hat V}_{\pv',\pv}\}).\nn
\eer
In the last equation the  summation over the repeated index $k$ is understood.
Correspondingly, the  Keldysh collision kernel  acquires two contributions to order $\lambda_0^4$ and reads
\be
\label{keldysh_kernel}
\delta I_K\equiv -i\left[\TSigmaC_{EY},\TGC\right] = -i\left[\TSigmaC^{(0)}_{EY},\TGC\right]-i\left[\TSigmaC^{(1)}_{EY},\TGC\right].
\ee
The first term on the right hand side gives rise to the  EY spin relaxation \cite{Raimondi10},
and   contributes an extra relaxation channel on the right hand side of Eq.(\ref{bloch})
\ber
\label{EYrelaxation}
-\frac{1}{\tau_{EY}}
\begin{pmatrix}
  1    &  0  &0 \\
     0 & 1   &  0\\
    0  &   0 & d-2
\end{pmatrix}
({\bf S}-\chi {\boldsymbol\Delta})
\equiv -\hat \Gamma_{EY} ({\bf S}-\chi {\boldsymbol\Delta}) 
\eer
where we have introduced the dimensionality-dependent EY spin relaxation rate given by
\be
\label{EY_relaxation}
\frac{1}{\tau_{EY}}=\frac{4 (d-1)}{d^2}\frac{1}{\tau}\left( \frac{\lambda_0p_F}{2}\right)^4.
\ee
In the above $d=2,3$ is the dimensionality of the space where particles move. 
The $d=2$ case corresponds to the 2-dimensional electron gas case, where we have concentrated our attention until now.
The $z$ component of the spin is a constant of the motion and does not undergo relaxation in this case. 
However, the peculiarity of the new term we are going to derive appears also, and more remarkably, in the $d=3$ case.
For this reason we keep the dependence on the dimensionality from now on.
 
The Keldysh $(K)$ component of the second term in (\ref{keldysh_kernel}) reads
\ber
\delta I_K^K &=&
-i\left(\TG^R-\TG^A\right)\TSigma_{EY}^{(1), K}\nn\\
& -&i\left(\TSigma^{(1), R}_{EY}\TG^K-\TG^K\TSigma^{(1), A}_{EY}\right)
\nn\\
&\equiv&
\delta I^{(1)} + \delta I^{(2)},
\eer
having used that $\TG^{R,A}\sim\sigma^0$. 
In order to obtain the Bloch equation we need to sum over the momentum as done  for obtaining the continuity equation (\ref{continuity}). The summation over momentum of the  Boltzmann collision integral\footnote{The Boltzmann collision integral is the Keldysh collision intgeral (\ref{keldysh_kernel}) integrated over the energy} is obtained as
\be
\sum_{\bp}\inten\delta I_K^K\equiv \Delta I^{(1)} + \Delta  I^{(2)}.
\ee
By replacing $\TG^R$, $\TG^A$ and $\TG^K$ with the expressions (\ref{cov_gr}-\ref{cov_gk}), one obtains 
\ber
\Delta I^{(1)} &=&\frac{n_i\pi}{2}
\sum_{\bp'\bp}\delta(\epsilon_{\bp}-\epsilon_{\bp'})\nn\\
&&
\Big(\partial_{p'_k}{\hat V}_{\pv,\pv'}\{\Acal_k,(1-2f_{\pv'})\}{\hat{V}}_{\pv',\pv}\nn\\
&+&\partial_{p_k}\{\Acal_k,{\hat V}_{\pv,\pv'}(1-2f_{\pv'}){\hat{V}}_{\pv',\pv}\Big) ,
\label{collision1}
\eer
and
\ber
\Delta I^{(2)} &=&-\frac{n_i\pi}{2}
\sum_{\bp'\bp}\delta(\epsilon_{\bp}-\epsilon_{\bp'})\nn\\
&&\frac{1}{2}\Big\{
(\partial_{p'_k}{\hat V}_{\pv,\pv'}2\Acal_k{\hat V}_{\pv',\pv}\nn\\
&+&\partial_{p_k}\{\Acal_k,{\hat V}_{\pv,\pv'}{\hat V}_{\pv',\pv}\}),(1-2f_{\pv})\Big\}.\nn\\
\label{collision2}
\eer
In both of the above equations, the first term, after the delta function,  has been obtained by an integration by parts with respect to the momentum $\bp'$. As a result, the derivatives with respect to $p'_k$ and $p_k$ act on the $\hat V_{\pv,\pv'}$ factors only.
In Eq.(\ref{collision1}) the dependence on the directions of the momentum $\pv$ is restricted to the $\hat V_{\pv,\pv'}$ factors only, so that one can perform at once the integration over the solid angle of $\pv$ and then take the derivative with respect to $\pv'$. Appendix \ref{appa} provides some useful identities  (see Eqs.(\ref{useful_identity}-\ref{useful_identity_2}))  on how to carry out these operations.
Notice also that the second term in round brackets of Eq.(\ref{collision1}) vanishes, because the derivative with respect to
$\pv$ yields a linear dependence on $\pv$ so that the solid angle integral gives zero. 

By reasoning in the same way, one sees that the first term in round brackets within the anticommutator  of Eq.(\ref{collision2}) also vanishes.
In the  second term one can make at once the integration over the solid angle of $\pv'$, again by using the results of Appendix \ref{appa}.  As a result, after 
 working out the Pauli algebra, one gets
\begin{eqnarray}
\label{collisionfull}
\Delta I &=&
\pi n_iv_0^2 \left(\frac{\lambda_0}{2}\right)^4\frac{d-1}{d}\sum_{\bp'\bp}\delta(\epsilon_{\bp}-\epsilon_{\bp'})
p^2 \left[ \frac{d-2}{d-1} \right.
\nn\\
&&  \left( \sigma^k\anticomm{e\Acal_k}{ p'_lf(\epsilon_{\bp'})}\sigma^l\right.\nn\\
&+&\left.
\sigma^l\anticomm{e\Acal_k}{ p'_l f(\epsilon_{\bp'})}\sigma^k\right) \nn\\
&-&\left.  \sigma^i\anticomm{e\Acal_k}{ p'_k f(\epsilon_{\bp'})}\sigma^i\right.\nn\\
&+&\left.  2p'^2  \anticomm{e\Acal_k}{ {p}_k f(\epsilon_{\bp}) }\right]. \label{eq_73}
\end{eqnarray} 
In Eq.(\ref{eq_73}) the summation over repeated indices runs over $x,y,z$ for $d=3$.  
For $d=2$, the last two lines of Eq.(\ref{eq_73}) survive and only the $i=z$ term remains.
Then the sum over momentum of the Boltzmann collision integral is
\footnote{{The corrections to the charge collision integral $\Delta I^0$,
relevant for the reciprocal SGE/IEE case in e.g. spin pumping setups, 
require considering higher-order terms and are discussed in \cite{toelle2016}.}}
\ber
\Delta I^a &=& 
\frac{1}{2}{\rm Tr}[\sigma^a\Delta I]
\nn\\
&=&
\frac{1}{\tau} \left(\frac{\lambda_0}{2}\right)^4p_F^2 \frac{(d-1)}{2d}\sum_{\bp} f^0(\epsilon_{\bp}) p_i \\
&\times &\Big(      e\Acal^a_i +\frac{d-2}{d-1}(e\Acal^i_a+e\Acal^n_n\delta_{ai}). \Big)\nn
\eer
This is {\it zero} as long as $f^0(\epsilon_{\bp})$ is isotropic, which is the case in a homogeneous system at equilibrium.
Things change as soon as an electric field is switched on and carriers have a finite drift velocity ${\bf v}=-e\tau {\bf E}/m$. We then have the spin generation torque due to the interplay of intrinsic and extrinsic SOCs ${\cal T}^a_{ext, sg}\equiv \Delta I^a$
\ber
{\cal T}^a_{ext, sg}& =&-\frac{N_0}{2\tau_{EY}}\Big(e\Acal^a_i
+\frac{d-2}{d-1}(e\Acal^i_a+e\Acal^n_n\delta_{ai}) \Big) v_i\nn\\
&=&{\cal C}_i^av_i\label{newtor}
\eer
where we have introduced the extrinsic SOC torque tensor ${\cal C}_i^a$. 
In $d=3$ it is instructive to represent this tensor as follows
\ber
{\cal C}_i^a&=&-\frac{e N_0}{2\tau_{EY}}\left[\Acal_n^n\delta_{ia}+\frac{3}{2}\left(\frac{1}{2}\left(\Acal_i^a+\Acal^i_a\right) -\frac{1}{3}\Acal_n^n\delta_{ia}\right) \right.\nn\\
&+&\left. \frac{1}{2}\left(\Acal_i^a-\Acal^i_a \right) \right]\label{3dtensor}
\eer
by separating explicitly all irreducible tensor parts - the unit, the traceless symmetric, and antisymmetric contributions.
Comparing this with the similar representation for the plain $\Acal^a_k$
we see the the symmetric (``Dresselhaus'') part
has a contribution 3 times as large relative to the antisymmetric (``Rashba") part.
Hence Eq.~\eqref{3dtensor} shows that the value at which the spin polarization would like to relax to by EY processes has a form different from the SOC internal field defined in Eq.(\ref{Bgeneral}) due to DP processes. The latter has the same structure as the first term in the brackets of Eq.(\ref{newtor}) but with an opposite sign. Three-dimensional motion adds an entirely new term to the internal SOC field induced by the electric field. Although when going to $d=3$ the linear DSOC may not be appropriate anymore, the overall message is that the interplay of extrinsic SOC and SU(2) intrinsic SOC is extremely rich. The exploration of the consequences of this are however beyond the scope of the present paper.

 For $d=2$ only the first term in the brackets of Eq.(\ref{newtor}) survives, so that,
by considering  $a=y$ for RSOC $(e\Acal_x^y=-2m\alpha)$, we have 
\ber
{\cal T}^y_{ext, sg}
&=&-\frac{1}{\tau_{EY}} N_0 \alpha m v_x.
\label{TokaCa1}
\eer	
Hence, the spin generation torque due to the interplay of RSOC and extrinsic SOC has the opposite sign with respect to the corresponding term originating by the Dyakonov-Perel precessional relaxation ${\cal T}^y_{int, sg}=1/{\tau_{DP}}(N_0 \alpha m v_x)$.
We name this new term the Elliott-Yafet torque (EYT).

\color{black}
We can then write the Bloch equation in the final form 
\begin{equation}
\label{bloch66}
\partial_t  {\bf S}=-{\boldsymbol\Omega}\times {\bf S}- (\hat \Gamma_{DP}+\hat \Gamma_{EY}) \left({\bf S}-
\chi{\boldsymbol \Delta}\right)+
{\mathbfcal T}_{sg},
\end{equation}
where   the spin generation torque ${\mathbfcal T}_{sg}$, in the presence of extrinsic SOC is given by  
\be
\label{total_torque}
{\mathbfcal T}_{sg}={\mathbfcal T}_{int, sg}+\delta {\mathbfcal T}_{int, sg}+{\mathbfcal T}_{ext, sg}, 
\ee
where
\ber
{\mathbfcal T}_{int, sg}&=&\hat\Gamma_{DP} \chi{\bf B}\label{int_torque}\\
\delta {\mathbfcal T}_{int, sg}&=& \frac{\theta_{SH}^{ext}}{\theta_{SH}^{int}}\hat\Gamma_{DP} \chi{\bf B}\label{ext_torque_SHE}\\
{\mathbfcal T}_{ext, sg}&=&-\hat\Gamma_{EY} \chi{\bf B}.
\label{ext_torque_EY}
\eer
Hence, the extrinsic SOC yields two additional spin generation torques (\ref{ext_torque_SHE}) and
(\ref{ext_torque_EY}) associated to spin Hall effect (to order  $\lambda_0^2$) and
Elliott-Yafet processes (to order  $\lambda_0^4$), respectively. 
The  second torque has the same form but opposite sign of the intrinsic torque, 
indicating that the EY spin-relaxation is detrimental to the ISGE/EE as anticipated
in Sec. \ref{sec_statement}.
The Bloch equations (\ref{bloch66}) together with the expressions of the various torques
(\ref{total_torque}-\ref{ext_torque_EY}), the DP ($\hat \Gamma_{DP}$) and EY (
$\hat\Gamma_{EY}$) spin relaxation matrices (\ref{DPmatrix}) and (\ref{EYrelaxation})
and the definition of the total magnetic field ${\boldsymbol\Omega}$ (\ref{effmagx}-\ref{effmagy}) are the main result of this paper.
In accordance with the experimental observations of Ref.~\cite{Norman2014},
Eq.~\eqref{bloch66} shows that, in general, the static non-equilibrium spin polarization will not be aligned along 
the internal effective magnetic field ${\boldsymbol \Omega}$.

\section{Conclusions}
\label{sec_conclusions}

In this paper we have considered the phenomenon of spin orientation by current by analyzing the interplay of 
intrinsic (Rashba and Dresselhaus) and extrinsic spin-orbit coupling.
We have derived the Bloch equation governing the spin dynamics by identifying the various relaxation and spin generation torques. 
Whereas in the presence of purely intrinsic spin-orbit coupling 
the spin polarization follows the internal effective magnetic field, this no longer happens when the extrinsic spin-orbit is present. 
The precise relation between the spin polarization and the Rashba-Dresselhaus internal field depends on the relative magnitude 
of the Dyakonov-Perel and Elliott-Yafet spin relaxation rates, as well as on the spin Hall angle in the system. 
These observations may be very useful in analyzing existing experiments on the ISGE/EE,
and in suggesting new ones.

\acknowledgements{CG acknowlegdes finacial support from the DFG (SFB 689). The work of IVT is supported by Spanish Ministerio de Economia y Competitividad (MINECO) through Project No. FIS2016-79464-P and by the 'Grupos Consolidados UPV/EHU del Gobierno Vasco' (Grant No. IT578-13). GV was supported  by NSF Grant No. DMR-1406568.}  

\appendix
\section{The covariant Green function in terms of Wilson lines}
\label{appb}
The {\it locally} covariant Green function is defined as
\be
 \check {\tilde G}(x_1,x_2)=U_{\Gamma}(x,x_1)\check G(x_1,x_2)U_{\Gamma}(x_2,x)
\label{appb_1}
\ee
where 
\be
U_{\Gamma}(x,x_1)={\cal P}\exp \left(- i \int_{x_1}^x eA^{\mu}(y){\rm d}y_{\mu}\right).
\label{appb_2}
\ee
The line integral of the gauge field is referred to as the Wilson line. In Eq.(\ref{appb_2}) ${\cal P}$ is a path-ordering operator. Since the Wilson line transforms covariantly under a gauge transformation $O(x)$
 \be
 U_{\Gamma}(x,x_1)\rightarrow O(x)U_{\Gamma}(x,x_1) O^{\dagger}(x_1),
 \label{appb_3}
 \ee
one easily sees that the covariant Green function $\check{\tilde G}$ transforms in a locally covariant way
\be
\check{\tilde G}(x_1,x_2)\rightarrow O(x) \check{\tilde G}(x_1,x_2) O^{\dagger}(x).
\label{appb_4}
\ee
To lowest order in the gauge field, one may expand the exponential of the Wilson line and, after Fourier transforming with respect ot the relative coordinate,  obtain Eqs.(\ref{shift1}-\ref{shift2}) of the main text. 

The Wilson line is unitary in the sense that
\be
U_{\Gamma}(x,x_1) U_{\Gamma}(x_1,x)=1.
\label{appb_5}
\ee
The unitarity of the  Wilson line allows to express the covariant transformation of a convolution product  of non covariant objects in terms of the convolution of the covariant transformed objects. In particular, the covariant transformation of the Keldysh collision integral gives 
\ber
&&U_{\Gamma}(x,x_1) \left[ \check\Sigma (x_1,x_3)\overset{\otimes}{,}\check G(x_3,x_2)\right] U_{\Gamma}(x_2,x)\nn\\
&=&
\left[\check{\tilde \Sigma} (x_1,x_3)\overset{\otimes}{,}\check{\tilde G}(x_3,x_2) \right]
\label{appb_6}
\eer
after using the unitarity of the Wilson line by inserting 
$$U_{\Gamma}(x_3,x) U_{\Gamma}(x,x_3)=1$$
 between the self-energy and the Green function.

\section{An identity concerning angular integration}\label{appa}
In the text we need to perform the  integration over the solid angle of  $\pv$ 
\be
\int \left( \frac{\sin (\theta_{\pv}) {\rm d}\theta_{\pv} }{2}\right)^{d-2}\frac{{\rm d}\phi_{\pv} }{2\pi}  \hat V_{\pv,\pv'} \dots \hat V_{\pv',\pv}
\equiv 
\langle \hat V_{\pv,\pv'} \dots \hat V_{\pv',\pv}\rangle.
\ee
In the above the dots  indicate any operator acting on the spin indices, but not depending on the momenta $\pv$ and $\pv'$.
By writing explicitly the cross products in the $\hat V_{\pv,\pv'}$ factors one has
\ber
&&-v_0^2\left(\frac{\lambda_0}{2}\right)^4\langle \sum_{ijklmn} \epsilon_{ijk}\epsilon_{lmn}p_ip'_j\sigma^k \dots p_lp'_m\sigma^n\rangle\nonumber \\
&=&-v_0^2\left(\frac{\lambda_0}{2}\right)^4 \sum_{ijklmn} \epsilon_{ijk}\epsilon_{lmn} \langle p_ip_l\rangle p'_jp'_m \sigma^k \dots \sigma^n\nonumber\\
&=&-v_0^2\left(\frac{\lambda_0}{2}\right)^4 \frac{p^2}{d}\sum_{ijklmn} \epsilon_{ijk}\epsilon_{lmn} \delta_{il}p'_jp'_m \sigma^k \dots \sigma^n\nonumber\\
&=&-v_0^2\left(\frac{\lambda_0}{2}\right)^4  \frac{p^2}{d}\left(p'^2 \sigma^i \dots \sigma^i \right.\nn\\
&-&\left. (d-2) \pv'\cdot {\boldsymbol\sigma} \dots \pv'\cdot {\boldsymbol\sigma} \right)\label{useful_identity}
\eer
where in $d=3$ it is understood a summation over $i=x,y,z$ and in $d=2$ $i=z$. If the dots are replaced by the identity in the spin space
\be
\langle \hat V_{\pv,\pv'} \dots \hat V_{\pv',\pv}\rangle =-v_0^2\left(\frac{\lambda_0}{2}\right)^4 \frac{2p^2p'^2}{d}\sigma^0.
\ee
Then the derivative with respect to $p'_k$ yields
\ber
&&\partial_{p'_k} \langle \hat V_{\pv,\pv'} \dots \hat V_{\pv',\pv}\rangle\nn\\
&=&-v_0^2\left(\frac{\lambda_0}{2}\right)^4  \frac{p^2}{d} \left( 2p'_k \sigma^i \dots \sigma^i \right.\nn\\
&-&\left.  \sigma^k \dots {\pv'}\cdot  {\boldsymbol\sigma} -{\pv'}\cdot  {\boldsymbol\sigma} \dots \sigma^k\right)
\label{useful_identity_2}
\eer

\end{document}